\documentclass[10pt,conference]{IEEEtran}
\usepackage{makecell}
\usepackage{cite}
\usepackage{amsmath,amssymb,amsfonts}
\usepackage{amsmath}
\usepackage{algorithm}
\usepackage{algpseudocode}
\usepackage{graphicx}
\usepackage{textcomp}
\usepackage{xcolor}
\usepackage{hyperref}
\usepackage{booktabs}

\def\BibTeX{{\rm B\kern-.05em{\sc i\kern-.025em b}\kern-.08em
    T\kern-.1667em\lower.7ex\hbox{E}\kern-.125emX}}

\begin{document}

\title{Towards a Transformer-Based Pre-trained Model for IoT Traffic Classification}

\author{\IEEEauthorblockN{Bruna Bazaluk\IEEEauthorrefmark{1}, Mosab Hamdan\IEEEauthorrefmark{2}, Mustafa Ghaleb\IEEEauthorrefmark{2}, Mohammed S. M. Gismalla\IEEEauthorrefmark{3}, \\Flavio S. Correa da Silva\IEEEauthorrefmark{1},
Daniel Macêdo Batista\IEEEauthorrefmark{1}}

\IEEEauthorblockA{\IEEEauthorrefmark{1}Department of Computer Science,
University of São Paulo, Brazil\\}

\IEEEauthorblockA{\IEEEauthorrefmark{2}Interdisciplinary Research Center for Intelligent Secure Systems, KFUPM, Saudi Arabia\\
\IEEEauthorblockA{\IEEEauthorrefmark{3}Center for Communication Systems and Sensing, KFUPM, Saudi Arabia\\}
Email:\{bazaluk, fcs, batista\}@ime.usp.br, \{mosab.mohamed, mustafa.ghaleb, mohammed.gismalla\}@kfupm.edu.sa}
}

\maketitle

\begin{abstract}
  The classification of IoT traffic is important to improve the efficiency and security of IoT-based networks. As the state-of-the-art classification methods are based on Deep Learning, most of the current results require a large amount of data to be trained. Thereby, in real-life situations, where there is a scarce amount of IoT traffic data, the models would not perform so well. Consequently, these models underperform outside their initial training conditions and fail to capture the complex characteristics of network traffic, rendering them inefficient and unreliable in real-world applications. In this paper, we propose IoT Traffic Classification Transformer (ITCT), a novel approach that utilizes the state-of-the-art transformer-based model named TabTransformer. ITCT, which is pre-trained on a large labeled MQTT-based IoT traffic dataset and may be fine-tuned with a small set of labeled data, showed promising results in various traffic classification tasks. Our experiments demonstrated that the ITCT model significantly outperforms existing models, achieving an overall accuracy of 82\%. To support reproducibility and collaborative development, all associated code has been made publicly available.
\end{abstract}


\section{Introduction}

IoT traffic classification stands as a crucial pillar in the realm of network management. It is an effective tool to improve the efficiency and security of the Internet of Things (IoT) ecosystem. The ability to classify IoT traffic accurately empowers Internet Service Providers (ISPs) to furnish high-quality services to network users, thereby ensuring optimal performance, security, and resource allocation \cite{velichkovska2023machine}. 

Conventional traffic classification techniques, which distinguish various network services based on basic traffic characteristics like communication protocol and port number, are increasingly inadequate due to the complexity and changeability of contemporary traffic~\cite{taylor2016appscanner, mohammed2020edge}. To overcome this challenge, numerous studies have employed machine learning (ML) algorithms for traffic classification using statistical features~\cite{oliveira2022stacked, shen2020optimizing, malekghaini2023deep}. Nonetheless, these methods depend heavily on expert judgment for selecting specific features, and even seemingly minor statistical features can significantly influence the effectiveness of the analysis.

Current methods for IoT traffic classification predominantly lean on deep learning (DL) algorithms~\cite{fitic, learning_based, multi}. One limitation is that these approaches hinge upon the availability of substantial volumes of labeled traffic data to construct traffic-level fingerprinting models. Thus, a significant challenge surfaces when handling specific classes of IoT devices that, during their operation, generate only limited labeled traffic.

Many ML and DL models nowadays use the transformers architecture~\cite{transformers}. Even though transformers were initially built to tackle problems in  Natural Language Processing (NLP), the successful utilisation of this architecture has become widespread, including problems related to e.g. image classification and tabular data analysis.
This way, working with transformers can be a good and innovative idea to solve the IoT traffic classification problem. In fact, there are some works that already used transformers to classify IoT traffic, but some do not use specific IoT traffic datasets \cite{blockchain} and others focus on specific type of networks~\cite{smart-home}.

This work introduces a novel IoT Traffic Classification Transformer (ITCT) model, based on TabTransformer~\cite{tabtransformer}, for classifying IoT traffic. The ITCT model harnesses the benefits of Transformers, such as efficient learning through an attention mechanism, the ability to generalize using extensive datasets, and proficiency in sequence learning problems, which is pertinent since network packet data inherently forms a sequence. Our model can learn network dynamics from packet traces, and the obtained results from the experiments are promising: ITCT demonstrates a remarkable ability to generalize across various prediction tasks and environments. Having been evaluated with generic datasets, the model employs Transformers to learn contextual embeddings of categorical features effectively. We have pre-trained this model using an MQTT-based IoT traffic dataset~\cite{dataset}, enabling others to further fine-tune it with their own data, regardless of the dataset's size. The results indicate the ITCT transformer's potential to achieve commendable evaluation metrics. For example, one of our proposed models attained an overall accuracy of 82\%, a similar performance to other classifiers tested. Furthermore, to promote transparency and reproducibility, all developed code is made publicly available\footnote{\url{https://github.com/brunabazaluk/tabtransformer_iot_attacks}}.


This paper is organized as follows: we discuss some related works in Section~\ref{sec:rw}. Section~\ref{sec:prop} explain out the methodology and experiments. Finally, Section~\ref{sec:experiments} and Section~\ref{sec:conclusions} present the experimental result and discussions and the conclusions, respectively. 

\section{Related Work} \label{sec:rw}

There are some studies regarding IoT traffic classification that try to solve the problem of the small quantity of data. One of the most recent 
uses feature comparison to classify the packets~\cite{learning_based}. The model consists of 10 different feature comparators, each being a neural network with a different combination of the following layers: Convolutional, batch normalization, ReLU, and Max Pooling. The Precision values vary between 80.34\% and 99.04\%, the Recall values fluctuate between 80.69\% and 99.01\%, and the F1-measure values span from 80.47\% to 99.01\% in two different datasets. However, this work suffers from computational costs.

One of the most recent studies regarding IoT traffic classification 
obtained great results~\cite{multi}. In the paper, the authors describe MAFFIT (multi-perspective feature approach to few-shot classification of IoT traffic), a model based on BiLSTMs (Bidirectional Long Short-Term Memory Recurrent Neural Networks). The main modules of this model is the feature encoder, feature comparator and finetune optimizer. The purpose of feature encoder is to further extract the high-dimensional features of the packet length sequences and the packet byte sequence. The feature comparator does the final predictions and the finetune optimizer finetunes the parameters of the training phase in order to achieve better results. This model achieved the best results in the area so far, in terms of accuracy, using IoT traffic datasets. 
Another work 
has a slightly different approach~\cite{agnostic}. The main idea is to transform the information in the packets into images and use them as input to a meta-learning model based on Convolutional Neural Networks. The authors used five different network traffic datasets, which were general and not necessarily from IoT traffic,  and selected a few samples from each. 
The dataset called FSIDS-IoT was made to be used on few-shot models. 
The only measure described in the paper is the accuracy of their model, which ranged from 73.81\% to 92.19\%. However, this proposal also suffers from computational costs. 

In~\cite{smart-home} the authors proposed securing a smart home with a Transformer-Based IoT intrusion detection system. They created a model to classify 
the traffic combining network traffic and telemetry from the house's sensors data and training a transformer-based model with these combined data. This model achieved an accuracy of 98.39\%. However, this work  
may not perform well in general cases.

Our work focuses on using transformers to classify IoT traffic. Our approach is using TabTransformer~\cite{tabtransformer}. This architecture was proposed by the end of 2020. It uses the encoder part of the transformer architecture~\cite{transformers} to process the categorical features and concatenates them with a normalization of the continuous features; this concatenation is then passed through a multi-layer perceptron to predict the specified class.

\begin{figure*}[htb]
		\centering
        \includegraphics[width=0.9\textwidth]{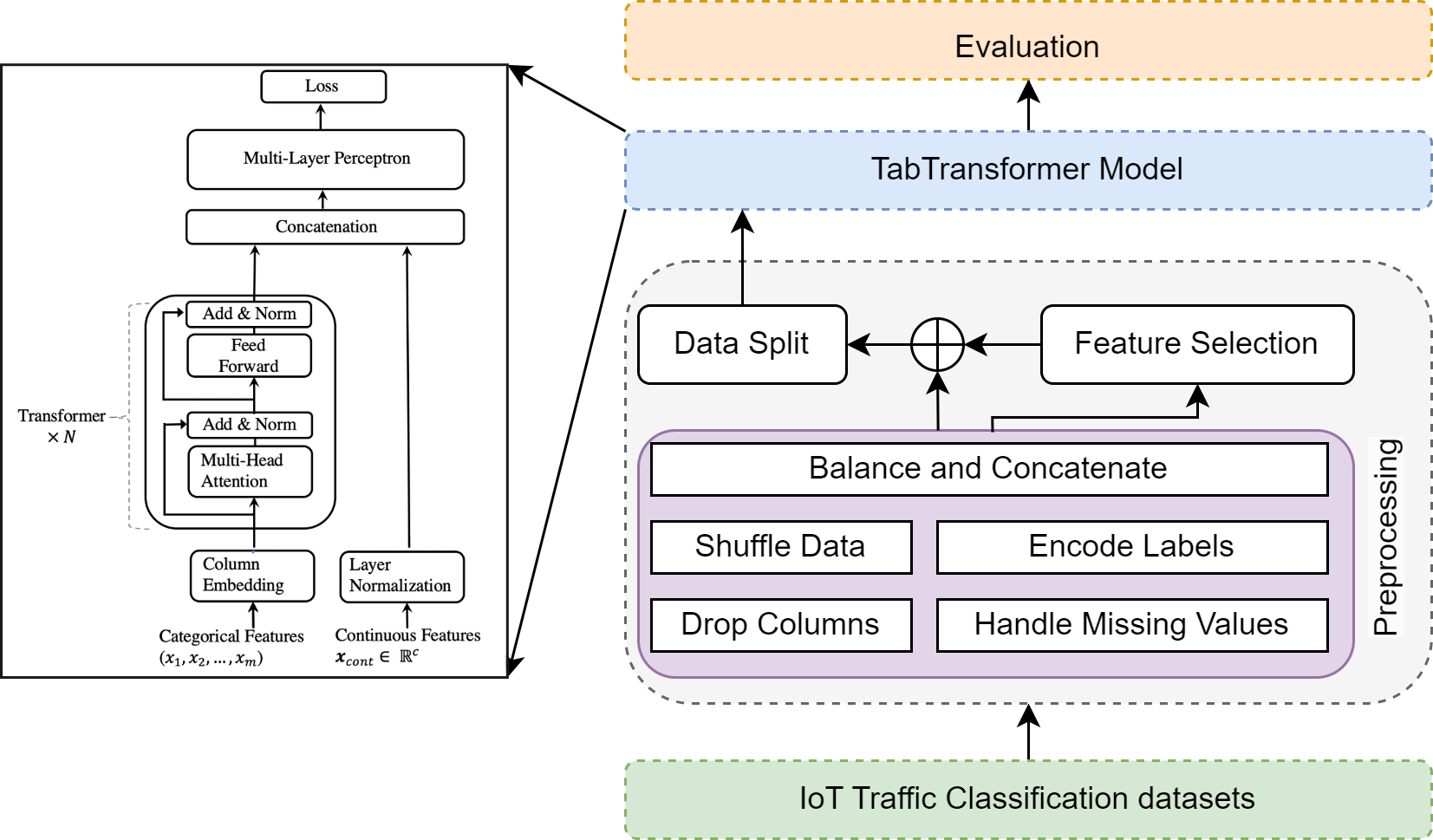}
	\caption{IoT Traffic Classification Transformer Model}
	\label{fig:layer}
\end{figure*}

\section{A New Transformer Model for IoT Traffic Classification}
\label{sec:prop}

Our proposed IoT Traffic Classification Transformer architecture, referred to as ITCT Transformer, is based on TabTransformer\cite{tabtransformer}, which comprises a column embedding layer, a stack of $N$ Transformer layers, and a multilayer perceptron (MLP). Each Transformer layer, as introduced by ~\cite{vaswani2017attention, tabtransformer}, consists of a multi-head self-attention layer followed by a position-wise feed-forward layer. The architecture of the IoT Traffic Classification Transformer is depicted in Figure \ref{fig:layer}.

\subsection{TabTransformer\cite{tabtransformer}}
Given a feature-target pair \((x, y)\), where \(x \equiv \{x_{\text{cat}}, x_{\text{cont}}\}\), \(x_{\text{cat}}\) represents all categorical features and \(x_{\text{cont}} \in \mathbb{R}^c\) denotes all of the continuous features. Each categorical feature \(x_i\) in \(x_{\text{cat}} \equiv \{x_1, x_2, \dots, x_m\}\) is embedded into a parametric embedding of dimension \(d\) through Column Embedding. The set of embeddings for all categorical features is denoted as \(E_\phi(x_{\text{cat}}) = \{e_{\phi_1}(x_1), \dots, e_{\phi_m}(x_m)\}\). These embeddings are then fed into the first layer of the Transformer. The output from the final Transformer layer, \(f_\theta\), transforms these parametric embeddings into contextual embeddings \(\{h_1, \dots, h_m\}\) where \(h_i \in \mathbb{R}^d\). These contextual embeddings, concatenated with continuous features \(x_{\text{cont}}\), form a vector of dimension \((d \times m + c)\), which is then inputted into an MLP, \(g_\psi\), to predict the target \(y\). The loss function \(L(x, y)\) is defined as 
\begin{equation}
L(x, y) \equiv H(g_\psi(f_\theta(E_\phi(x_{\text{cat}})), x_{\text{cont}}), y),
\end{equation}
where \(H\) is the cross-entropy for classification tasks and mean square error for regression tasks. This function aims to minimize the prediction error and learn all the Transformer parameters in an end-to-end manner.

For the Transformer, a multi-head self-attention mechanism is used, involving parametric matrices Key ($K$), Query ($Q$), and Value ($V$). The attention head transforms each input embedding into a contextual one through the attention mechanism defined as 
\begin{equation}
\text{Attention}(K, Q, V) = A \cdot V,
\end{equation}
where \( A = \text{softmax}((Q K^T)/\sqrt{K}) \). For column embedding, each categorical feature \(i\) has an embedding lookup table \(e_{\phi_i}(\cdot)\), with \(e_{\phi_i}(j) = [c_{\phi_i}, w_{\phi_{ij}}]\) where \(c_{\phi_i} \in \mathbb{R}\), \(w_{\phi_{ij}} \in \mathbb{R}^{d-1}\). Unique identifiers \(c_{\phi_i} \in \mathbb{R}\) distinguish classes in column \(i\) from those in other columns. The embeddings are pre-trained in a supervised manner using labeled examples and further refined through fine-tuning with labeled data, with the loss defined in Equation (1). For scenarios with limited labeled examples, pre-training procedures like Masked Language Modeling (MLM) and Replaced Token Detection (RTD) are employed.

\subsection{ITCT (IoT Traffic Classification Transformer)}
As detailed in Algorithm \ref{alg:Transformer_for_IoT}, the Transformer for IoT Traffic Classification model employs a systematic approach to categorize MQTT-IoT traffic data effectively. The process commences with a thorough data preprocessing phase, where numerical features are normalized and categorical features are suitably encoded. A critical step in this phase is the application of feature selection techniques, ensuring that the model focuses on the most pivotal attributes of the data. By the way, the pre-processing was made for each file separately and then they were concatenated in one dataset. The pipeline to be followed is to preprocess the dataset, make the feature selection, and train and test ITCT Transformer model with the selected features on the preprocessed dataset.  Subsequently, the algorithm proceeds to initialize a Transformer model, specifically designed and tuned to align with the nuanced characteristics of the MQTT-IoT dataset. Training this model involves carefully optimizing various parameters, guided by a predetermined loss function, optimizer, and critical performance metrics. Post-training, the model undergoes a rigorous evaluation on an independent test dataset, a crucial step to verify its robustness and effectiveness. In essence, the Transformer for IoT Traffic Classification stands as a comprehensive and potent tool, able to dissect and classify MQTT-IoT traffic data, enhancing performance and ensuring its practical viability in many real-world applications.

\begin{algorithm}
\footnotesize
\caption{Transformer for IoT Traffic Classification\label{alg:Transformer_for_IoT}}
\begin{algorithmic}[1]
\Require IoT traffic dataset, Number of epochs ($epochs$) for training, Batch size ($batch\_size$) for training
\Ensure Trained Transformer model, Performance metrics on the validation set

\Function{Preprocess\_Data}{$data$}
    \State Load the IoT traffic dataset
    \State Normalize numerical features in the dataset
    \For{each categorical feature $f$ in $data$}
        \State Apply appropriate encoding to $f$ (e.g., one-hot encoding,
        \State embedding)
    \EndFor
    \If{Feature Selection is applied}
        \State Apply feature selection techniques to focus on the most
        \State informative features
    \EndIf
    \State Split $data$ into training ($train\_data$), testing ($test\_data$), 
    \State and validation ($val\_data$) sets
    \State \Return $train\_data$, $test\_data$, $val\_data$
\EndFunction

\Function{Initialize\_Model}{$input\_features$, $num\_classes$}
    \State Define Transformer model architecture with $input\_features$ and 
    \State $num\_classes$
    \State Compile the model with loss function, optimizer, and performance 
    \State metrics
    \State \Return the initialized model
\EndFunction

\Function{Train\_Model}{$model$, $train\_data$, $val\_data$, $epochs$, $batch\_size$}
    \State Train the model using $train\_data$ with parameters $epochs$ and 
    \State $batch\_size$
    \State Validate the model using $val\_data$ after each epoch
    \State Optionally, use callbacks for early stopping/learning rate adjustment
    \State \Return training history $history$
\EndFunction

\Function{Evaluate\_Model}{$model$, $test\_data$}
    \State Evaluate the trained model on $test\_data$
    \State Calculate performance metrics: accuracy, AUC, precision, recall
    \State \Return evaluation results $results$
\EndFunction

\State $train\_data$, $test\_data$, $val\_data$ $\gets$ \Call{Preprocess\_Data}{IoT traffic dataset}
\State $model$ $\gets$ \Call{Initialize\_Model}{input features from $train\_data$, number of classes}
\State $history$ $\gets$ \Call{Train\_Model}{$model$, $train\_data$, $val\_data$, $epochs$, $batch\_size$}
\State $results$ $\gets$ \Call{Evaluate\_Model}{$model$, $test\_data$}
\State Print $results$
\end{algorithmic}
\end{algorithm}



\section{Experiments and Discussion} \label{sec:experiments}

To verify the effectiveness of ITCT, our proposed model, we
implemented a set of experiments detailed as follows.

\subsection{Dataset}

We chose an open IoT traffic dataset, \textit{MQTT Internet of Things Intrusion Detection Dataset} (MQTT-IoT-IDS2020)~\cite{dataset}. The two main reasons for our choice are: 1) This dataset focuses on the MQTT protocol, which is one of the most used protocols in IoT communication; 2) this is a large dataset with more than 1GB, and as our idea is to develop a pre-trained model that can be fine-tuned with only a few samples, a large dataset is needed.
We use the \texttt{.csv} version of the dataset, which is already tabular. This dataset was built under a simulated MQTT network. Its architecture contains twelve sensors, a broker, a simulated camera, and an attacker. The five captured scenarios are: aggressive scan, UDP scan, Sparta SSH brute-force, MQTT brute-force attack, and normal operation, as shown in Table~\ref{tab:mqtt_dataset}. 

\begin{table}[htb]
\renewcommand{\arraystretch}{1.3}
\caption{Characteristics of the MQTT Dataset}
\footnotesize
\label{tab:mqtt_dataset}
\centering
\begin{tabular}{lcc}
\toprule
Category & \makecell{Normal\\Operation} & Attacks \\
\midrule
Scana & 70,768 & 40,624 \\
Scansu & 210,819 & 22,436 \\
Sparta & 947,177 & 19,728,963 \\
MQTT Bruteforce & 32,164 & 10,013,152 \\
Normal & 1,056,230 & 0 \\
\bottomrule
\end{tabular}
\end{table}

The dataset is composed of 5 different files, 4 of them have samples from both attacks and normal operation, and one has only normal operation data. During the pre-processing, we balance each of the 4 first files to guarantee they have the same quantity of the two classes. We used the remaining attacks' samples from the files that have more attacks than normal operation samples, together with the samples of the last file, which has only the normal operation samples, so that this last file is also balanced.\\
Our approach to deal with missing values is to calculate the mean of the non-NaN values of each column and replace NaN with this mean.\\
Therefore, the preprocessing consists of: normalizing numerical features, dealing with NaN values, tokenizing the values and words to create the vocabulary to be used by our ITCT Transformer model, doing the feature selection and splitting the data between train, test and validation.

\subsection{IoT Traffic Data Preparation and Preprocessing}

After the preprocessing, we make the feature selection using Sci-kit Learn's Tree-based estimator \cite{sklearn} in order to separate the columns to be considered in the training process. From a total of 25 features, only 5 were selected: \textit{ttl, ip\_len, tcp\_flag\_push, mqtt\_message\_type, and mqtt\_message\_length}. Additionally, we include one more feature, namely, \textit{protocol}. 

We separate 80\% of the samples for the training, 10\% for validation and 10\% for testing. Then, we use TensorFlow's keras\footnote{\url{https://keras.io/examples/structured_data/tabtransformer}} to build the ITCT model. The first step is to define the vocabulary, which was defined using the tokens found on the categorical features of the data, tokenize it and pass it trough the embedding layer. Then, we define the optimizer to be used, in this case we use AdamW \cite{keras}, which is the default keras' optimizer and the most recommended for transformers-based networks, and the loss function, which is Binary Categorical Entropy, because we are working with only 2 classes. After setting the layers and parameters, the model is compiled.

Most of the parameters were the same as keras' TabTransformer webpage suggested, others were changed to fit our specific problem. This model is then trained with the selected dataset. To avoid overfitting, we use a validation set during the training process, that helps the model to know when to stop the training. The last 10\% of the data is used for testing. In the test, we use the trained model to classify the samples. The metrics we use are accuracy, precision, F1-Score, and recall.


\subsection{Experimental Settings}

\subsubsection{Parameters of the proposed ITCT model} 
In the implementation of our model, careful consideration is given to the selection and tuning of hyperparameters, ensuring optimal performance and efficient learning. These are shown on Table~\ref{tab:hyperparameters}.

\begin{table}[htb]
\renewcommand{\arraystretch}{1.3}
\caption{Hyperparameters of the Model}
\label{tab:hyperparameters}
\footnotesize
\centering
\begin{tabular}{lc}
\toprule
Hyperparameter & Value \\
\midrule
Learning Rate & 0.001 \\
Weight Decay & 0.0001 \\
Dropout Rate & 0.2 \\
Batch Size & 265 \\
Number of Epochs & 20 \\
Number of Transformer Blocks & 4 \\
Number of Attention Heads & 4 \\
Embedding Dimensions & 16 \\
MLP Hidden Units Factors & [2, 1] \\
\bottomrule
\end{tabular}
\end{table}

\normalsize

\subsubsection{Implementation}
The analytical procedures were implemented using Python version 3.7.10, with Keras and TensorFlow serving as the primary frameworks for deep learning tasks. For the purpose of replicability and to facilitate further research, we have provided all the experiments detailed in this section in a publicly accessible Python notebook \footnote{\url{https://github.com/brunabazaluk/tabtransformer_iot_attacks}}.

\begin{table*}
\renewcommand{\arraystretch}{1.3}
\caption{IoT Traffic Classification Transformer Approach in Various Configurations}
\label{tab:tab_transformer}
\begin{centering}
\centering
\begin{tabular}{lccc}
\toprule
\textbf{Metrics} & \textbf{Experiment 1 (w/FE \&  w/Callback)} & \textbf{Experiment 2 (w/o FE \& w/o Callback)} & \textbf{Experiment 3 (w/o FE \& w/Callback)} \\
\midrule
Accuracy(\%) & 77.00 & \textbf{82.00} & 81.63 \\
Precision & 0.84 & \textbf{0.87} & 0.82 \\
Recall & 0.77 & \textbf{0.82} & \textbf{0.82} \\
F1-Score & 0.81 & \textbf{0.84} & 0.82 \\
AUC ROC Score & 0.7651 & \textbf{0.8174} & 0.8163 \\
Training Time (seconds) & 1,009.84 & 1,226.46 & \textbf{363.21} \\
Inference Time (seconds) & \textbf{2.14} & 2.56 & 2.51 \\
Total Model Weights & \textbf{21,231} & 26,209 & 26,209 \\
\bottomrule
\end{tabular}
\par\end{centering}
\end{table*}

\begin{table*}
\renewcommand{\arraystretch}{1.3}
\caption{Comprehensive Comparison of Proposed and Existing Approaches}
\label{tab:comparison}
\begin{centering}
\centering
\begin{tabular}{lcccc}
\toprule
\textbf{Model} & \textbf{Accuracy (\%)} & \textbf{Precision} & \textbf{Recall} & \textbf{F1-measure} \\ \hline
FESTIC\cite{learning_based} & - & 0.80 - \textbf{0.99} & \textbf{0.81} - \textbf{0.99} & 0.80 - \textbf{0.99} \\ 
MAML-CNN\cite{agnostic} & 74 - \textbf{92} & - & - & -\\ 
\textbf{Proposed ITCT} & \textbf{77} - 82 & \textbf{0.82} - 0.87 & 0.77 - 0.82 & \textbf{0.81} - 0.84 \\ \hline
\end{tabular}
\par\end{centering}
\end{table*}

\subsection{Experimental Results and Analysis}

In the comparison of
ITCT Transformer configurations for IoT Traffic Classification model, distinct variations in performance are evident. Table~\ref{tab:tab_transformer} presents the experimental results.

\subsubsection{Performance metrics analysis}
In our analysis, Experiment 2 without Feature Extraction \& selection and Callbacks (w/o FE \& w/o Callback) has the highest accuracy (\textbf{82\%}) and precision (\textbf{0.87}), showcasing its effectiveness in correctly classifying IoT traffic. The recall rate stands at \textbf{0.82} for both Experiments 2 and 3. Notably, Experiment 2 without Feature Extraction \& selection and Callbacks (w/o FE \& w/o Callback) achieves the highest F1-Score (\textbf{0.84}), reflecting a balanced measure of precision and recall and asserting its robustness in scenarios where discerning between false positives and false negatives is critical.

The \textbf{AUC ROC Score} further underscores the distinction of Experiment 2 with a leading score of \textbf{0.8174}, attesting its capability in distinguishing between classes under varying threshold settings.

\subsubsection{Computational efficiency}
\textbf{Training Time:} Experiment 3 completes its training phase in \textbf{363.21 seconds}, significantly underpinning its counterparts. This expeditious training is paramount in environments where timely model updates are crucial or computational resources are scarce.

\textbf{Inference Time:} The swiftest inference is registered in Experiment 1 (\textbf{2.14 seconds}). Fast inference times are indispensable in real-time applications where prompt decision-making is imperative, such as instantaneous network threat detection within IoT infrastructures.

\textbf{Total Model Weights:} A notable reduction in model complexity is observed in Experiment 1, presenting only \textbf{21,231 weights}, in contrast to the \textbf{26,209 weights} in Experiments 2 and 3. This reduction not only curtails memory consumption but also accelerates inference times, making these models particularly well-suited for deployment in environments with limited computational resources. 

\subsubsection{Impact and implications}
The impact of \textbf{Feature  Selection \& Extraction and Callbacks} is pronounced, with Experiment 3 demonstrating significant reductions in model complexity and training duration. However, it is important to note that the highest performance in metrics like accuracy and precision is attained in Experiment 2 (w/o FE \& w/o Callback), suggesting that feature reduction could potentially omit vital information. On the other hand, the function loss value of 0.3721 for our model in experiment 2, incorporating both feature extraction and callback 
mechanisms, reflects its capability to minimize prediction errors effectively. Conversely, experiment 3, which employs callbacks (w/o FE \&  w/Callback), demonstrates a more refined model performance with a reduced loss value of 0.2179, indicating a superior fit to the training data.
\subsection{Comparison:} In Table~\ref{tab:comparison} we can see the comparison between the before cited methods and our work. Although our metrics are not the best, they are close to other methods and have the highest lower bounds. Furthermore, as discussed before, these other works suffer from computational costs.

We can see there is balance between model complexity and performance. While more streamlined models are favored for their computational efficiency, it is imperative to ensure that this simplification does not degrade the model's predictive strength.

Considering real-world applicability, especially in scenarios characterized by extensive data streams and limited computational resources, the choice of model configuration necessitates a balance between precision and operational constraints. Experiment 3, with its high F1-Score, and reduced training duration, is a great solution where both efficiency and accuracy are taken into consideration.

In conclusion, this analysis accentuates the trade-offs between accuracy, computational efficiency, and model complexity. The selection of a model configuration should be guided by the specific demands and limitations of the application environment, with a consideration of how feature extraction and callbacks influence the model's capacity to generalize and perform consistently. These insights are instrumental in refining IoT traffic classification models to align with diverse operational requirements.

\section{Conclusion and Future Work}
\label{sec:conclusions}

In this study, we introduce a novel IoT Traffic Classification model (ITCT), an approach based on the TabTransformer model, specifically designed for the classification of IoT traffic. The ITCT model showcases a good performance in traffic classification tasks compared to other existing IoT traffic classifiers\cite{fitic}\cite{multi}\cite{agnostic}, by integrating a traffic representation matrix and packet-level attention mechanisms. Notably, our model achieves great accuracy and precision, particularly in configurations without feature extraction and callbacks, demonstrating its effectiveness in correctly classifying IoT traffic. However, our findings also emphasize the importance of maintaining a balance between model complexity and performance, highlighting that simplified models can achieve substantial computational efficiency without significantly compromising predictive capabilities.

An interesting advantage our model has over the before cited methods is that we already trained ITCT with a large dataset\cite{dataset}, so, to use it in real life, users may fine-tune it with their own data to get better results for their specific environment.

In future work, we plan to make our model available at HuggingFace\footnote{\url{https://huggingface.co/}}, so that it will be easier to be fine-tuned by users. We also aim to improve the model's adaptability to various IoT scenarios and network environments, training it with different datasets.





\section*{Acknowledgment}

This research is part of the INCT of the Future Internet for Smart Cities funded by CNPq proc. 465446/2014-0, Coordena\c{c}\~ao de Aperfei\c{c}oamento de Pessoal de N\'ivel Superior – Brasil (CAPES) – Finance Code 001, FAPESP proc. 14/50937-1, and FAPESP proc. 15/24485-9. It is also part of the FAPESP proc. 21/06995-0. Furthermore, the authors Mosab Hamdan and Mustafa Ghaleb would like to acknowledge the support provided by the Interdisciplinary Research Center for Intelligent Secure Systems at KFUPM. Moreover, the author Mohammed S. M. Gismalla expresses appreciation for the support received from the Center for Communication Systems and Sensing at KFUPM.

\bibliographystyle{IEEEtran}
\bibliography{bib.bib}

\end{document}